%Paper: 9110043
%From: mermin@helios.TN.CORNELL.EDU (N. David Mermin)
%Date: Tue, 15 Oct 91 13:49:42 EDT

\baselineskip=20pt                      % Double spacing
\magnification=1200
\font\title = cmbx10 scaled 1440        % For big title

\def\and{{\rm and\ }}

\def\L{$\{\,\,$}
\def\R{$\,\,\}$}
\def\+{+}

\def\h{\fr1/2}
\centerline{{\title Copernican Crystallography}}
\bigskip
\centerline{N. David Mermin}
\centerline{Laboratory of Atomic and Solid State Physics}
\centerline{Cornell University, Ithaca, NY 14853-2501}

\bigskip {\it Abstract.\/} Redundancies are pointed out in the widely used
extension of the crystallographic concept of Bravais class to quasiperiodic
materials.  Such pitfalls can be avoided by abandoning the obsolete paradigm
that bases ordinary crystallography on microscopic periodicity.  The
broadening of ordinary crystallography to include quasiperiodic materials is
accomplished by defining the point group in terms of indistinguishable (as
opposed to identical) densities.

\bigskip

A periodic material (crystal) is characterized by a lattice of vectors [1]
specifying the translations that leave its density unchanged.  A century and
a half ago Frankenheim classified such lattices by their symmetry, counting
15 types.  A few years later Bravais pointed out that two of the Frankenheim
classes contained identical lattices, and today the edifice of
crystallography rests on a foundation of 14 Bravais classes [2].

Within the past two decades the discovery of many quasiperiodic materials
(displacively modulated crystals, substitutionally modulated crystals,
incommensurate intergrowth compounds, quasicrystals) has stimulated efforts
to extend the crystallographic classification scheme to include these novel
structures.  Roughly speaking quasiperiodic materials have the property ---
reminiscent of but weaker than periodicity --- that subregions of arbitrary
size can be found reproduced elsewhere in the material at distances of the
order of that size.  This notion is made precise by the Fourier space
definition of quasiperiodicity, which gives a simple and natural expression
of the close connection between periodic and quasiperiodic materials.
Densities of either type of material are superpositions of plane waves whose
wave vectors can be expressed as a lattice of integral linear combinations
of $3+d$ primitive wave vectors that span a three dimensional space and are
linearly independent over the integers.  A material is periodic if $d=0$
and quasiperiodic if $d>0$.  To emphasize that the vectors in such lattices
are wave vectors rather than translations, one may refer to them as
reciprocal lattices.

Although originally viewed as containing lattices of translations, the
Bravais classes of periodic materials can equally well be regarded as
classes of reciprocal lattices.  Since quasiperiodic materials have no
3-dimensional translational symmetry but continue to be described by a
lattice of wave vectors, it is only in Fourier space that the concept of
Bravais class can directly be applied to them.  The first attempt at such a
classification, using the less direct superspace formalism described below,
was made over a decade ago by Janner, Janssen, and de Wolff (JJdW) [3,4] for
the simplest quasiperiodic materials with crystallographic point group
symmetries.  Such materials can be described by a reciprocal lattice with
$d=1$ for 6 of the 7 crystals systems; for the 7th (cubic) system the
minimum $d$ is 3.  The JJdW catalog lists 24 (3+1) Bravais classes, and 14
(3+3) cubic Bravais classes.

The narrow purpose of this Letter is to point out that 8 pairs of the (3+1)
JJdW Bravais classes are identical --- {\it i.e.\/} each of the two classes
contains exactly the same lattices of 3-dimensional wave vectors --- as are
3 pairs and one trio of the (3+3) cubic Bravais classes.  The specific pairs
and the trio are listed in Tables 1 and 2.  Those engaged in
crystallographic studies of incommensurately modulated materials can readily
confirm these identifications by simply working out the general form of the
lattices of ordinary 3--dimensional wave vectors belonging to each JJdW
Bravais class.  A direct derivation in 3-dimensional Fourier space of these
16 (3+1) and 9 (3+3) cubic Bravais classes is given elsewhere [5].

My broader purpose is to comment on the reasons behind this redundancy of
description [6].  These and other anomalies in the existing generalization of
crystallography to quasi\-per\-iod\-ic materials are unlikely to disappear
until crystallographers abandon the venerable but outdated enshrinement of
periodicity as the {\it sine qua non\/} of their taxonomy.  By the time
quasiperiodic materials were discovered the view that crystallography is
limited to the classification of periodic materials was so entrenched that
the extension to quasiperiodic materials was achieved only by expressing
them as 3--dimensional sections of materials periodic in more than three
dimensions, to which the higher dimensional crystallography of periodic
materials could then be applied.  By liberating crystallography from
its historic reliance on periodicity, one can avoid climbing up into
superspace in search of periodicity for the hazardous purpose of coming back
down with a bag full of categories, simply by taking a step sideways from
3--dimensional position space into 3--dimensional Fourier space.  In Fourier
space, as already noted, the distinction between periodic and quasiperiodic
materials is elementary, and the fundamental concept of a Bravais class of
lattices can be trivially extended from the periodic to the quasiperiodic
case [7] without ever leaving 3 dimensions.

The virtues of Fourier space, even as the venue for the traditional
crystallography of periodic materials, were celebrated by A.~Bienenstock and
P.~P.~Ewald [8] three decades ago.  They pointed out that the 230
crystallographic categories of Sch\"onflies, Fedorov, and Barlow could be
derived simply and efficiently in Fourier space as classes of phase relations
between density Fourier coefficients at wave vectors related by point--group
operations.  Quasiperiodic materials not then having attracted serious
attention, Bienenstock and Ewald presented their method only as a more
powerful approach to the ordinary crystallography of periodic materials.
Their Fourier space classification scheme can, however, be directly derived
without any appeal to periodicity in 3 or any other number of dimensions as
the natural way to classify the broader class of quasiperiodic materials
[9--11]. Since Fourier space offers an unorthodox but more effective route to
the ordinary crystallography of periodic materials, since Fourier space
provides the simplest definition of quasiperiodicity, and since the Fourier
space route to crystallography applies equally well to both periodic and
quasiperiodic materials, the single advantage of ascending to superspace in
search of a classification scheme based on periodicity, is that it relieves
one of having to take a radical new look at the foundations of ordinary
crystallography [12].

The key to a reformulation of crystallography that does not rely on
periodicity, and perhaps the most important inducement for working in
Fourier space [13], emerges from the concept of indistinguishable --- as
opposed to identical --- densities.  Two densities are indistinguishable if
their positionally averaged $n$--point autocorrelation functions are the
same for all $n$ --- {\it i.e.\/} if any substructure on any scale that
occurs in one occurs in the other with the same frequency.  Two periodic
densities are so strongly constrained by the condition of periodicity that
they can be indistinguishable only if they differ by at most a translation,
but two quasiperiodic densities can be indistinguishable even when they are
not so simply related.

The concept of indistinguishable densities resolves a puzzle about
quasiperiodic materials (such as 5-fold Penrose tilings).  Many of them
clearly appear to possess certain real--space point--group symmetries even
though no origin can be found about which a point group operation takes the
density into itself at arbitrary distances.  (Although this is not often
emphasized, they therefore lack strict rotational as well as translational
symmetry.)  How is this to be understood?  The puzzle is resolved by
redefining the point group of a quasiperiodic material to be the set of
operations from $O(3)$ that take the density into one that is not {\it
identical to\/}, but merely {\it indistinguishable from\/} the original.
With the relaxation of identity to indistinguishability, such point group
operations become strictly valid, and in fact apply about any origin
whatever.

Should a material happen to be periodic, indistinguishability reduces to
identity to within a translation; one is then led to extend the point group
to a space group which includes the translations that combine with point
group operations to leave the density identical to what it was.  Should the
material be quasiperiodic, however, translations cease to be relevant.  The
role of the space group in classifying such materials is played by the point
group, supplemented by an elementary consequence of the use of
indistinguishability in the definition of the point group.
Indistinguishability assumes a very simple form in Fourier space:  two
densities are indistinguishable if the products of their Fourier
coefficients over any set of wave vectors summing to zero always agree.
This leads straight\-forwardly to a classification scheme, valid for
periodic or quasiperiodic materials, that is based on point group symmetry
and the phase relations between density Fourier coefficients at symmetry
related wave vectors [9--11].  For periodic materials the resulting
categories can also be regarded as subgroups of the real--space Euclidean
group (the crystallographic space groups) because of the simple form
indistinguishability assumes in the presence of periodicity.  For
quasiperiodic materials, however, and for a powerful and broader approach to
periodic materials as well, one is better off building the classification
scheme out of the point group of operations that take the density into
something indistinguishable, and the phase relations between Fourier
coefficients related by operations from that point group [14].

Tables 1 and 2 provide a simple example of the importance of staying in
3-dimensions.  The redundancies in the JJdW catalog correspond to the
different (3+0) crystallographic sublattices that can be found within any
3--dimensional lattice from the given (3+$d$) Bravais class [15]. The
particular way in which the lattices in two such JJdW classes are embedded in
superspace obscures the fact that both classes contain identical sets of
ordinary 3--dimensional wave-vectors.  Information about the different (3+0)
sublattices of the (3+$d$) lattices in a given Bravais class is quite useful
in applying the scheme to the diffraction patterns of those quasiperiodic
materials where such (3+0) sublattices can be associated with a
crystallographic lattice of main reflections and the remaining wave vectors,
with weaker satellite peaks.  But to count as distinct Bravais classes these
different ways in which one and the same class of $(3+d)$ lattices can be
manifest in particular materials is to abandon the view of a Bravais class
as a class of lattices of 3--dimensional wave vectors, that serve as a
template for a diffraction pattern independent of the particular intensities
of the Bragg peaks that are actually observed [16].  By building such
distinctions into the concept of Bravais class one unnecessarily creates
different crystallographies for displacively modulated crystals (where the
distinctions are relevant) and quasicrystals or incommensurate intergrowth
compounds (where they are not); one obscures relations such as that between
the 3 Bravais classes of icosahedral quasicrystals and the 3 (not 4) (3+3)
cubic Bravais classes with tetrahedral symmetry, noted in the caption of
Table 2; and when one comes to compute the classes of phase relations (which
give the space groups in the periodic case and are currently displayed in
the form of superspace groups in the quasiperiodic case), by failing to
recognize that some Bravais classes contain identical 3--dimensional
lattices of wave vectors one ends up calculating and tabulating the same
information more than once.

If the Sun were the only thing of interest in the heavens, it would be
foolish not to regard it as moving around the Earth.  Because this view
became firmly entrenched, generations of astronomers had to learn about
epicycles to account for the motions of the planets.  While it was wrenching
to shift to a heliocentric perspective, the resulting simplification in the
more broadly applied scheme more than made up for the pain of abandoning the
Ptolemaic view.  When all materials of interest were periodic, a
crystallography based on periodicity grew and thrived.  Epicycles appeared,
in the form of superspace groups, when the scheme was extended to
quasiperiodic materials without abandoning its conceptual reliance on
periodicity.  While unwilling to burn for it at the stake, I would like to
suggest that others could spare themselves significant pain by abandoning
Ptolemaic crystallography and learning how to classify both periodic and
quasiperiodic materials, not by ascending to superspace but by resting the
foundations of crystallography on the concept of a point group of operations
that change the density into something indistinguishable.

{\it Acknowledgments.\/} My appreciation for the power of a theory of
symmetry based on the notion of indistinguishable densities developed in the
course of collaborations with Daniel Rokhsar, David Wright, Jason Ho, and
David Rabson.  I am indebted to Ron Lifshitz for joining me in a detailed
analysis of the JJdW Bravais classes, and to Sander van Smaalen, whose
thoughtful response to that analysis led me to wonder whether
crystallography might not be in need of a paradigm shift.  Veit Elser, Chris
Henley, and Michael Widom commented helpfully, with varying degrees of
approbation or disapprobation, on earlier versions of the manuscript.  This
research is supported by the National Science Foundation, Grant No. DMR
8920979.
\bigskip
\centerline{{\bf References}}
\medskip

\item{[1]} A set of vectors is a lattice if {\bf v}$-${\bf w} is in the set
whenever {\bf v} and {\bf w} are.

\item{[2]}The contributions of Frankenheim (1842) and Bravais (1848) are
described in A.~E.~H. Tutton, {\it Crystallography and Practical Crystal
Measurement\/}, MacMillan, London, 1922.

\item{[3]} P.~M.~de Wolff, T.~Janssen, A.~Janner, Acta Cryst. A{\bf 37},
625 (1981).

\item{[4]} A.~Janner, T.~Janssen, and P.~M.~de Wolff, Acta Cryst A{\bf
39}, 658, 667, 671 (1983).

\item{[5]} N.~D.~Mermin and R.~Lifshitz, submitted to Acta Crystallographica
A.

\item{[6]} I stress that the redundancy is there, independent of my
diagnosis of the underlying problem.  I comment further on the redundancy in
the context of that diagnosis, at the end of this Letter.

\item{[7]} Ref. [5] gives a geometrical definition for when two $(3+d)$
lattices are in the same Bravais class, but it is not relevant to the
problem with the JJdW Bravais classes noted here, that Bravais classes
counted as distinct in their scheme sometimes contain {\it identical\/}
lattices of 3-dimensional wave-vectors.

\item{[8]} A.~Bienenstock, and P.~ P.~ Ewald, Acta Cryst.{\bf 15},
1253 (1962).

\item{[9]} D.~S.~Rokhsar, D.~C.~Wright, and N.~D.~Mermin, Acta Cryst. A{\bf
44}, 197 (1988).

\item{[10]} D.~A.~Rabson, N.~D.~Mermin, D.~S.~Rokhsar, and D.~C.~Wright,
Revs. Mod.  Phys. {\bf 63}, 699 (1991).

\item{[11]} N.~D.~Mermin, to appear in Revs. Mod.  Phys. {\bf 64}, January,
1992.

\item{[12]} Superspace can, of course, be used advantageously for other
purposes, such as building models that suggest where the atoms might be in
real 3-space.  See, for example, V. Elser and C. L. Henley, Phys. Rev. Lett.
{\bf 55}, 2883 (1985).

\item{[13]} X-ray diffractionists should need no such inducement.  Since
they see the microscopic material directly in Fourier space, what could be
more natural than to formulate the crystallographic categories there as well?

\item{[14]} These phase relations play no role in the particular application
made here, which addresses only the concept of Bravais class.  I mention
them (a) because a simplification of the concept of Bravais class that left
one unable to formulate finer details of the classification scheme would be
worthless, and (b) because they provide, through their connection to the
real space concept of indistinguishable densities, one of the most important
conceptual reasons for working in Fourier space.

\item{[15]} This is particularly clear for the pairs of JJdW Bravais classes
associated with the classes I have numbered 20, 21, and 22 in Table 2.

\item{[16]} All but a finite number of the peaks will, of course, not be
observed at all.  The lattice itself should be viewed as a mathematical
abstraction from the diffraction pattern: the set of all integral linear
combinations of wave vectors determined by the observed Bragg peaks.
 \bigskip
\centerline{{\bf Table Captions}}
\medskip

{\sl Table 1.} The 16 (3+1) Bravais classes for the non-cubic crystal
systems.  They contain lattices with the point group symmetry of the crystal
system, made up of vectors that are integral linear combinations of 4
integrally independent vectors.  As the nomenclature in the first column is
intended to indicate, 14 of the 16 Bravais classes contain lattices that are
simply given by all the sums of all pairs of vectors, one from a lattice in
a coventional crystallographic Bravais class, and the other, an arbitrary
integral multiple of a single additional vector.  The second and third
columns list the names and numbers of the corresponding JJdW Bravais class
(or classes).  When two JJdW classes appear on the same line, they describe
identical lattices of wave-vectors, but viewed from the perspective of one
or the other of two different 3+0 crystallographic sublattices.  Redundant
JJdW classes are listed within braces ($\{\ \}$).  The correspondence between
the Bravais class designations in the first and second columns should be
obvious in the 14 simple cases.  The JJdW labels characterize the lattices
in terms of a crystallographic sublattice (specified by the capital letter)
to each vector of which is added arbitrary integral linear combinations of
the vector listed in parentheses and all its images under the point group of
the crystal system.  In the first column the designations $I^*$ and $F^*$
refer to what are conventionally called the orthorhombic $F$ and $I$ Bravais
classes, as is appropriate for a classification scheme based in Fourier
space.

\medskip

{\sl Table 2.} The 9 (3+3) cubic Bravais classes. The first 6 have full cubic
symmetry; the last 3, only tetrahedral symmetry.  Lattices in all 6 of the
full cubic Bravais classes contain sums of all pairs of vectors taken from
two incommensurate (3+0) cubic lattices, each of these crystallographic
lattices belonging to any of the 3 crystallographic cubic Bravais classes,
as the nomenclature in the first column indicates.)  The JJdW symbols and
numbers are given in the second column and (when a Bravais class occurs
under more than one name in their catalog) in the third and fourth.  The
convention behind their nomenclature is as described in the caption of Table
1.  The JJdW symbols in the second column for the lattices with full cubic
symmetry clearly reveal their relation to the designations in the first
column.  Redundant JJdW symbols in the third and fourth columns are given in
braces ($\{\ \}$) to emphasize that they do not describe additional classes
of lattices.  As in Table 1, $F^*$ is synonymous with but preferable to $I$
(and similarly for $I^*$ and $F$) since the Bravais classes contain lattices
of vectors in Fourier space.  If one ignores superspace and considers
wave-vectors in ordinary 3-dimensional Fourier space, it should be evident
that the Bravais classes with full cubic symmetry described by the redundant
JJdW symbols in the third column contain lattices of wave-vectors identical
to those in the Bravais classes described by the symbols in the second.
Establishing this for the symbol in the fourth column requires a small
amount of analysis.  The remaining three cubic (3+3) Bravais classes contain
lattices with only tetrahedral symmetry.  Qua\-si\-crystal\-lographers might
note that the three icosahedral $F^*$, $P$, and $I^*$ lattices are nothing
but the (3+3) cubic lattices $T_0$, $T_1$, and $T_2$ (respectively), with
the ratio of the two incommensurate length scales set at a special value
that raises the point group symmetry from tetrahedral to icosahedral.

\vfil
\eject
\centerline{{\bf TABLE 1}}
\vskip0.7truein
\def\a{\alpha}
\def\b{\beta}
\def\c{\gamma}
\def\fr#1/#2{{\textstyle{#1\over#2}}}
\def\h{\fr1/2}
\def\hf{\hfill}
\def\th{\fr1/3}
\def\tt{\thinspace\thinspace}
\def\n#1{\hf{\bf\ #1}}
\centerline{
\setbox\strutbox=\hbox{\vrule height12pt depth6.5pt width0pt}
\vbox{\offinterlineskip
\hrule
\hrule
\hrule
\halign{
        \strut \vrule\hfil\tt #\tt\hfil
              &\vrule\hfil\tt #\tt\hfil
              &\vrule\hfil\tt #\tt\hfil\vrule\cr
\multispan3\vrule\strut\hfill {\bf Triclinic} \hfill\vrule\strut\cr
\noalign{\hrule}
$P+1$ \n1   & $P(\a\b\c)$\hf {\it 1}           & \cr
\noalign{\hrule}
\noalign{\hrule}
\noalign{\hrule}
\multispan3\vrule\strut\hfill {\bf Monoclinic} \hfill\vrule\strut\cr
\noalign{\hrule}
$P+1_{ab}$ \n2 & $P2/m(\a\b0)$ \hf{\it 2}  &                   \cr
$B+1_{ab}$ \n3 & $B2/m(\a\b0)$ \hf{\it 4}  & \L $P2/m(\a\b\h)$ \hf{\it 3}\R
\cr
$P+1_c$   \n4  & $P2/m(00\c)$  \hf{\it 5}  &                            \cr
$C+1_c$   \n5  & $B2/m(00\c)$  \hf{\it 7}  & \L $P2/m(\h0\c)$  \hf{\it 6}\R
\cr
$M$       \n6  & $B2/m(0\h\c)$ \hf{\it 8}  &                            \cr
\noalign{\hrule}
\noalign{\hrule}
\noalign{\hrule}
\multispan3\vrule\strut\hfill {\bf Orthorhombic} \hfill\vrule\strut\cr
\noalign{\hrule}
$P+1$     \n7  & $Pmmm(00\c)$ \hf{\it 9}   &                            \cr
$I^*+1$   \n8  & $Fmmm(00\c)$ \hf{\it 17}   & \L $Pmmm(\h\h\c)$\hf{\it 11}\R
\cr
$F^*+1$   \n9  & $Immm(00\c)$ \hf{\it 12}   & \L $Cmmm(10\c)$ \hf{\it 14}\R
\cr
$C+1$     \n{10} & $Cmmm(00\c)$ \hf{\it 13}   &                            \cr
$A+1$     \n{11} & $Ammm(00\c)$ \hf{\it 15}  & \L $Pmmm(0\h\c)$ \hf{\it 10}\R
\cr
$O$       \n{12} & $Ammm(\h0\c)$ \hf{\it 16} & \L $Fmmm(10\c)$ \hf{\it 18}\R
\cr
\noalign{\hrule}
\noalign{\hrule}
\noalign{\hrule}
\multispan3\vrule\strut\hfill {\bf Tetragonal} \hfill\vrule\strut\cr
\noalign{\hrule}
$P+1$     \n{13} & $P4/mmm(00\c)$ \hf{\it 19} &                        \cr
$I+1$ \n{14} & $I4/mmm(00\c)$ \hf{\it 21} & \L $P4/mmm(\h\h\c)$ \hf{\it 20}\R
\cr
\noalign{\hrule}
\noalign{\hrule}
\noalign{\hrule}
\multispan3\vrule\strut\hfill {\bf Trigonal} \hfill\vrule\strut\cr
\noalign{\hrule}
$R+1$ \n{15} & $R\bar3m(00\c)$ \hf{\it 22} & \L $P\bar31m(\th\th\c)$
                              \hf {\it 23}\R  \cr
\noalign{\hrule}
\noalign{\hrule}
\noalign{\hrule}
\multispan3\vrule\strut\hfill {\bf Hexagonal} \hfill\vrule\strut\cr
\noalign{\hrule}
$P+1$ \n{16} & $P6/mmm(00\c)$ \hf {\it 24}  &
\cr
\noalign{\hrule}
\noalign{\hrule}
\noalign{\hrule}}}}
\vfil\eject

\vskip0.2truein
\bigskip
\def\a{\alpha}
\def\b{\beta}
\def\c{\gamma}
\def\h{\fr1/2}
\def\hf{\hfill}
\def\th{\fr1/3}
\def\tt{\thinspace\thinspace}
\centerline{{\bf TABLE 2}}
\vskip1truein
\centerline{
\setbox\strutbox=\hbox{\vrule height12pt depth6.5pt width0pt}
\vbox{\offinterlineskip
\hrule
\hrule
\hrule
\halign{
        \strut \vrule\hfil\tt #\tt\hfil
              &\vrule\hfil\tt #\tt\hfil
              &\vrule\hfil\tt #\tt\hfil
              &\vrule\hfil\tt #\tt\hfil\vrule\cr
\multispan4\vrule\strut\hfill {\bf Cubic} \hfill\vrule\strut\cr
\noalign{\hrule}
$P+P$    \n{17}        & $Pm3m(\a00)$   \hf {\it 208}             &&\cr
$I^* + I^*$  \n{18}    & $Fm3m(\a\a\a)$ \hf{\it 217}           &&\cr
$F^* + F^*$  \n{19}    & $Im3m(0\b\b)$ \hf{\it 213}           &&\cr
$P+I^* = I^*+P$ \n{20}  & $Pm3m(\a\a\a)$ \hf{\it 215} &
  \L $Fm3m(\a00)$ \hf{\it 211}\R &\cr
$P+F^* = F^*+P$ \n{21} & $Pm3m(0\b\b)$ \hf{\it 212} &
  \L $Im3m(\a00)$ \hf{\it 210}\R  &\cr
$I^* + F^*
=F^* + I^*$  \n{22}    & $Im3m(\a\a\a)$ \hf{\it 216}
                &\L $Fm3m(\a00)$   \hf{\it 214}\R
                &\L $Pm3m(\a\h\h)$ \hf{\it 209}\R       \cr
\noalign{\hrule}
\noalign{\hrule}
\noalign{\hrule}
\multispan4\vrule\strut\hfill {\bf Tetrahedral} \hfill\vrule\strut\cr
\noalign{\hrule}
$T_0$    \n{23}       & $Pm3(\h\b\b+\h)$ \hf {\it 206} & & \cr
$T_1$ =
    $T_0 + I^*$ \n{24}   & $Fm3(1\b\b+1)$   \hf {\it 207}
               & \L $Pm3(\a\h0)$     \hf {\it 204}\R  & \cr
$T_2 = T_0
 + I^* + I^*$ \n{25}  & $Fm3(\a10)$      \hf {\it 205} & & \cr
\noalign{\hrule}
\noalign{\hrule}
\noalign{\hrule}}}}
\bigskip
\bye